\documentclass[showpacs,aps,twocolumn,graphicx]{revtex4}
\usepackage{amsmath}
\usepackage{amscd}
\usepackage{graphicx}
\begin{document}

\title{Efficient polarization qubit transmission assisted by frequency degree of freedom}
\author{ Xi-Han Li\footnote{Corresponding author: xihanli@cqu.edu.cn}}
\address{Department of Physics, College of Physics, Chongqing University,
Chongqing 400044, China }
\date{\today }

\begin{abstract}
We present an efficient arbitrary polarization qubit transmission
scheme against channel noise by utilizing frequency degree of
freedom, which is more stable in transmission surroundings. The
information of quantum state is encoded in frequency state during
the transmission and transferred to polarization state later. Both
the fidelity of quantum state transmitted and the success
probability of this scheme are 1 in principle.
\end{abstract}

 \maketitle


The transmission of quantum states between participates is an
essential step in each quantum communication protocol. However, the
fidelity of a quantum state rapidly degrades because of the photon
loss and decoherence during the transmission, which will impact the
efficiency and the security of quantum communication consequently.
The coupling between quantum system and the environment, which is
called as noise, is a serious obstacle to a perfect communication.
The polarization degree of freedom (DOF) of photons are usually
chosen as information carries because of its maneuverability.
However, this DOF is incident to be influenced by thermal
fluctuation, vibration and imperfection of fibers, i.e, the channel
noise. Various methods had been proposed for rejecting or correcting
the errors caused by noise, one of which can be called as quantum
error rejection \cite{pla343, apl91, prl95, OC}.

In quantum error rejection schemes, there is an important assumption
that the variation of noise is slow. If several qubits transmit
through the channel simultaneously or close to each other, the
impact of noise is identical. With this hypothesis, arbitrary
quantum state can be transmitted probabilistically by making two
parts suffer from the same noise to interact and then reject errors
by postselection. In 2005, Kalamidas proposed a single-photon error
rejection protocol \cite{pla343}, which encodes the quantum states
in two time-bins and an uncorrupted state can be obtained at a
definite time of arrival. However, fast polarization modulator are
employed in this scheme, whose synchronization makes it difficult to
implement. Later, we presented a faithful qubit transmission scheme
with passive linear optics \cite{apl91}. Yamamoto \emph{et al}
proposed a qubit distribution scheme with ancillary qubit in 2005
\cite{prl95}. However, the success probability is a bit low. All of
these three protocols utilize the temporal DOF to encode the state
where losses will be introduce inevitably during the postselection
to rebuild the original state.

Recently, some other DOFs besides polarization attract much
attention, such as frequency, spatial mode, orbital angular
momentum, transverse spatial mode, and so on. Since operations in
one DOF do not perturb the others, photons with multi-DOF may
provide novel ways to realize quantum information process. Frequency
DOF has been used in a series of quantum information schemes because
of its stability
\cite{OC,freqkd1,freqkd2,freqkd3,freqkd4,fre-distri,fre-puri,free-puri2}.
We proposed an efficient qubit transmission scheme \cite{OC} with
frequency DOF whose success probability is four times of Ref.
\cite{prl95}.

In this letter, we present a new method for transmitting a single
photon polarization state against channel noise with frequency DOF.
The sender encodes the information of quantum state into frequency
DOF and the receiver converts the information back to the
polarization DOF and obtain the  uncorrupted state
deterministically.


Let us suppose that an arbitrary single-photon pure state to be
transmitted is written as
\begin{eqnarray}
\vert \psi \rangle =\alpha \vert H \rangle + \beta \vert V
\rangle,\,\,\, (\vert \alpha \vert^2 + \vert \beta \vert^2=1).
\end{eqnarray}
Here $\vert H \rangle$ and $\vert V \rangle$ represent the
horizontal and the vertical polarization states of a single photon,
respectively. And $\alpha$ and $\beta$ are two parameters which
carry the information of the state to be transmitted. The frequency
of this photon is prepared in $\omega_2$. Considering the two DOFs,
the whole quantum state to be transmitted can be written as
\begin{eqnarray}
\vert \psi \rangle =\alpha \vert H,\omega_2 \rangle + \beta \vert V,
\omega_2 \rangle.
\end{eqnarray}
Before the transmission, the sender encodes the state with the
encoder composed of a polarizing beam splitter (PBS), a beam
splitter (BS) and a half wave plate (HWP), shown in Fig.1. PBS
transmits the horizontal polarization mode $\vert H \rangle$ and
reflects the vertical one $\vert V \rangle$. In this way, the state
$\vert H \rangle$ propagates through the path 1 and the state $\vert
V \rangle$ goes though path 2. The two pathes have the same length.
The frequency shifter  (FS) in path 1 modulates the frequency from
$\omega_2$ to $\omega_1$ and the HWP in path 2 affects the
polarization modes as the bit-flip operation $\sigma_x$, i.e.,
$\vert H \rangle \leftrightarrow \vert V \rangle$.

\begin{figure}[!h]
\centering
\includegraphics[height=1.2in]{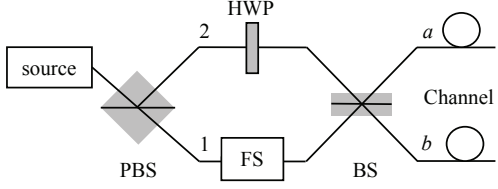}
\caption{The schematic diagram of the encoder. Path 1 and 2 have the same length.
The Sender encodes his qubit into a frequency superposition state and transmits it
to the receiver through two independent noisy channels.} \label{1}
\end{figure}

The initial state evolves in the encoder as follows,
\begin{eqnarray}
\vert \psi \rangle & =&\alpha \vert H,\omega_2 \rangle + \beta \vert V, \omega_2 \rangle\nonumber\\
&\xrightarrow{{\tiny \; PBS \;}}& \alpha \vert H,\omega_2 \rangle_1
+ \beta \vert V, \omega_2 \rangle_2\nonumber\\
&\xrightarrow[{\tiny \; HWP \;}]{{\tiny \; FS \;}}& \alpha \vert H,\omega_1 \rangle_1
+ \beta \vert H, \omega_2 \rangle_2\nonumber\\
&\xrightarrow{{\tiny \; BS \;}}& \frac{1}{\sqrt{2}}[ (\alpha \vert H,\omega_1 \rangle
+ i\beta \vert H, \omega_2 \rangle)_a \nonumber\\
&&\;\; + \;\; (i\alpha \vert H,\omega_1 \rangle + \beta \vert H,
\omega_2 \rangle)_b].
\end{eqnarray}
The subscripts 1 and 2 represent the two pathes of the
interferometer, and $a$ and $b$ represent the two output ports of BS
and two independent noisy channels. The coefficient $i$ comes from
the the phase shift aroused by the reflection of BS. From the last
two lines, we find the information of the initial state is encoded
into a frequency superposition state, which is immune to the channel
noise. And these two states in different noise channel have similar
form, we just analyze the situation of channel $a$ in detail below.

The channel noise can be expressed with a unitary transformation
\begin{eqnarray}
\vert H \rangle\rightarrow\delta \vert H \rangle + \eta \vert V \rangle,\,\, (\vert \delta \vert^2 + \vert \eta \vert^2=1)
\end{eqnarray}
where $\delta$ and $\eta$ are the parameters of the noise. The quantum state the receiver get from channel $a$ can be
written as
\begin{eqnarray}
&&\frac{1}{\sqrt{2}} (\alpha \vert H,\omega_1 \rangle
+ i\beta \vert H, \omega_2 \rangle)_a\xrightarrow{{\tiny \; noise \;}} \vert \psi \rangle_a\\\nonumber
&&=\frac{1}{\sqrt{2}} (\alpha \delta\vert H,\omega_1 \rangle +\alpha \eta \vert V,\omega_1 \rangle
+ i\beta \delta\vert H, \omega_2 \rangle +i\beta \eta \vert V, \omega_2 \rangle)
\end{eqnarray}

\begin{figure}[!h]
\centering
\includegraphics[height=1.2in]{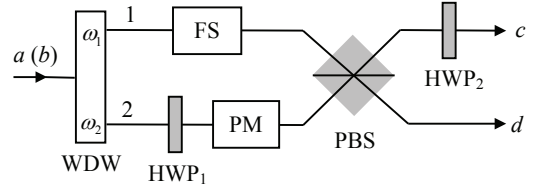}
\caption{The schematic diagram of the decoder. Wavelength division multiplexer (WDW) guides photon
to different spatial modes according to its frequency. Half wave plates  (HWP) act as a bit-flip
operation and the phase modulator (PM) adjusts the phase by $\pi/2$. The frequency shifter  (FS)
modulates the frequency so as to eliminate the frequency difference between different pathes.}
\end{figure}

The receiver uses a decoder to obtain the original polarization
state, shown in Fig.2. The wavelength division multiplexer (WDW) is
polarization independent and used to guide photons to different
spatial modes according to their frequencies. That is, frequency
state $\omega_1$  ($\omega_2$) goes to path 1  (2) according to the
figure. The FS in path 1 adjusts the frequency state from $\omega_1$
to $\omega_2$, so as to make the quantum states in these two pathes
have the same frequency. And the HWP in path 2 rotates the
horizontal polarization $\vert H \rangle$ to $\vert V \rangle$. The
phase modulator (PM) modulates the phase of quantum state passing
through path 2 as $\vert \psi \rangle \rightarrow e^{i\varphi}\vert
\psi \rangle$, where $\varphi=\pi/2$. The effect of the decoder can
be written as
\begin{eqnarray}
\vert \psi \rangle_a &\xrightarrow{{\tiny \; WDM \;}}&\frac{1}{\sqrt{2}} (\alpha \delta\vert H,\omega_1 \rangle_1
+\alpha \eta \vert V,\omega_1 \rangle_1 \nonumber\\&&+ i\beta \delta\vert H, \omega_2 \rangle_2
+i\beta \eta \vert V, \omega_2 \rangle_2)\nonumber\\
&\xrightarrow[{\tiny \; HWP_1 \;}]{{\tiny \; FS,PM \;}}&\frac{1}{\sqrt{2}} (\alpha \delta\vert H,\omega_2 \rangle_1
+\alpha \eta \vert V,\omega_2 \rangle_1 \nonumber\\&&+ \beta \delta\vert V, \omega_2 \rangle_2
+\beta \eta \vert H, \omega_2 \rangle_2)\nonumber\\
&\xrightarrow[{\tiny \; HWP_2 \;}]{{\tiny \; PBS
\;}}&\frac{1}{\sqrt{2}}[\delta (\alpha \vert H,\omega_2 \rangle +
\beta \vert V, \omega_2 \rangle)_d\nonumber\\&&+ \eta (\alpha \vert
H,\omega_2 \rangle  +\beta \vert V, \omega_2 \rangle)_c].
\end{eqnarray}
The subscripts 1 and 2 represent two pathes and $c, d$ represent two output ports of the decoder.
These two states in two different spatial modes have the same form with the initial one, retaining
the information of the quantum state. With the decoder set in channel $a$, the receiver can get the
initial state with success probability 1/2.
The situation of channel $b$ is similar. Suppose the noise is $\vert H \rangle\rightarrow \delta '
 \vert H \rangle + \eta ' \vert V \rangle$ in channel $b$, the state received with another decoder set in channel $b$ is
\begin{eqnarray}
\vert \psi \rangle_b &\xrightarrow{{\tiny \; decoder
\;}}&\frac{i}{\sqrt{2}}[\delta' (\alpha \vert H,\omega_2 \rangle +
\beta \vert V, \omega_2 \rangle)_d\nonumber\\&&+ \eta' (\alpha \vert
H,\omega_2 \rangle  + \beta \vert V, \omega_2 \rangle)_c],
\end{eqnarray}
which are also equal states with the original one. Therefore the
total success probability to transmit an arbitrary polarization
state is 1 in principle.


In the present scheme, the frequency DOF is introduced to defeat the channel
noise. Before the transmission, the secret information encoded in
polarization state was transferred to frequency DOF, which will be
converted back to original polarization state after transmission.
Therefore the modulation of frequency DOF is important in our
protocol. The FS, which can be carried out by acousto-optic
modulator (AOM) \cite{fbs,aom} or sum-frequency generation process
\cite{sfg}, is used to shift the frequency state of photons. And the
WDM works as a frequency beam splitter \cite{fre-distri,fre-puri},
which guides photons with different frequency to different spatial
modes, can also be substituted by a fiber bragg grating  (FBG)
\cite{freqkd3,freqkd4}. In a word, the operations required by our
scheme can be implemented with current technology.

In our scheme, two decoders are used because the quantum state
transmits through two independent noisy channels. In order to save
costs, only one decoder is required with the upgrade of the encoder,
as shown in Fig.3. The two states exit from two output ports of BS
are guided to the same noisy channel by means of an unbalance
interferometer. Pathes 3 and 4 have different length and a state
passing through path 3 will be delay by $\Delta T$.
\begin{figure}[!h]
\centering
\includegraphics[height=1.2in]{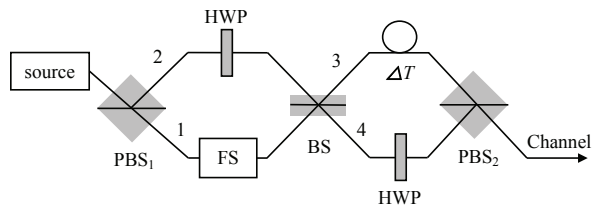}
\caption{The schematic diagram of the upgraded encoder.
The encoded quantum state is imported into one noisy channel by means of an
unbalance interferometer. During the transmission, the superposition of state is both in time and frequency.}
\end{figure}
Then the state launched into the noisy channel can be described as
\begin{eqnarray}
\frac{1}{\sqrt{2}}[ (\alpha \vert H,\omega_1 \rangle &+& i\beta \vert H, \omega_2 \rangle)_{t+\Delta T} \nonumber\\
+ (i\alpha \vert V,\omega_1 \rangle &+& \beta \vert V, \omega_2
\rangle)_t].
\end{eqnarray}
Here $t$ and $t+\Delta T$ denotes the arrival time, where $t$ is the arrival time of the state which did not be delayed.
Suppose the channel noise acts as
\begin{eqnarray}
\vert H \rangle\rightarrow \delta_1 \vert H \rangle +\eta_1 \vert V
\rangle,\,\,\, \vert V \rangle\rightarrow \delta_2 \vert H \rangle
+\eta_2 \vert V \rangle.
\end{eqnarray}
 Then state after the decoder becomes
\begin{eqnarray}
\frac{1}{\sqrt{2}}[\delta_1 (\alpha \vert H,\omega_2 \rangle &+& \beta \vert V, \omega_2 \rangle)_{d (t+\Delta T)}\nonumber\\
+ \eta_1 (\alpha \vert H,\omega_2 \rangle  &+&\beta \vert V, \omega_2 \rangle)_{c (t+\Delta T)}]\nonumber\\
+\frac{i}{\sqrt{2}}[\delta_2 (\alpha \vert H,\omega_2 \rangle &+& \beta \vert V, \omega_2 \rangle)_{d (t)}\nonumber\\
+ \eta_2 (\alpha \vert H,\omega_2 \rangle  &+& \beta \vert V,
\omega_2 \rangle)_{c (t)}].
\end{eqnarray}
From the expression we can find that the receiver can obtain the
quantum state in two different spatial modes $c$ and $d$ and in two
time slots $t+\Delta T$ and $t$. However, whenever and wherever the
quantum state arrives, it is always entirely consistent with the
original one. It means a perfect transmission of an initially
arbitrary polarization state through a noisy channel.

This scheme proposed an useful method to transmit polarization qubit against
channel noise. Then it will have good applications in one way quantum
communication protocols such as BB84 QKD protocol \cite{bb84}. The two parties can choose the two nonorthogonal bases, $\vert \pm
x\rangle=\frac{1}{\sqrt{2}}(\vert H \rangle \pm \vert V \rangle)$
and $\vert \pm y\rangle=\frac{1}{\sqrt{2}}(\vert H \rangle
\pm i\vert V \rangle)$, to realize the sharing of secret keys. With our encoder and decoder, the efficiency of quantum key distribution scheme is the same as the original one and the errors caused by the channel noise are eliminated.

In summary, we have proposed a single-photon error correction scheme
against collective noise with frequency DOF. In this scheme,
additional qubits is not required. The polarization state is encoded
in frequency DOF which is immune to noise and decrypted after
transmission. The success probability is 100\% in principle. As in
our scheme two pulses traveling in different time slots are not
interacted to get the uncorrupted state, the channel noise should
not be restricted as collective one. In addition, our scheme is not
only suitable for pure states but also for mixed ones, which will
have good applications in one-way quantum communication.


This work is supported by the Fundamental Research Funds for the
Central Universities Project under Grant No. CDJZR10100018.

\end{document}